\documentclass[reprint,superscriptaddress,aps,longbibliography]{revtex4-2}
\usepackage[utf8]{inputenc}
\usepackage[T1]{fontenc}
\usepackage{amsmath}
\usepackage{amssymb}
\usepackage{graphicx}
\usepackage[capitalize, nameinlink]{cleveref}
\usepackage{subfigure}
\graphicspath{{figures/}}
\usepackage{adjustbox}
\usepackage{booktabs}
\usepackage{multirow}

\usepackage{color}
\usepackage{rotating}
\usepackage{ulem}

\begin{document}

\title{Wave dynamics in a macroscopic square artificial spin ice}

\author{Lawrence A. Scafuri}
\author{Dmytro A. Bozhko}
\author{Ezio Iacocca}
\affiliation{Center for Magnetism and Magnetic Nanostructures, University of Colorado Colorado Springs, Colorado Springs, CO 80918, USA}

\date{\today}

\begin{abstract}
A macroscopic square artificial spin ice, or macro-ASI, is a collection of bar magnets placed in a square lattice arrangement. Each magnet is supported by hinges that allow their mechanical rotation. Previous investigations in these structures have shown ground-state configurations and driven dynamics similar to those of their nanosized counterparts. Here, we numerically investigate the impact of a defect, a Dirac string, on the driven dynamics. We find that waves quickly lose coherence by scattering between modes in this system. In addition, we observe a distinct mode associated with the isolated oscillation of the Dirac string. As expected from spatial localization, this resonant mode is under the band of propagating waves in the macro-ASI. The results are analogous to the development of low-frequency edge modes in nanoscopic spin ices in the presence of defects. Our results provide valuable insight into the physics of mechano-magnetic systems, demonstrating the existence of wave scattering and spatial confinement phenomena in macro-ASIs.
\end{abstract}

\maketitle

\section{Introduction}

Artificial spin ices (ASIs) are metamaterials composed of nanomagnets arranged in a geometrical fashion~\cite{Skjaervo2020}. Many different magnetic configurations are available due to their geometric frustration~\cite{Heyderman2013,Wang2006,Gilbert2014}, which has fueled an interest in their applications for magnonics~\cite{Gliga2020,Lendinez2019}, i.e., the control of dynamical excitations of magnetization order---magnons. Reconfigurable band structures have been numerically predicted in these metamaterials~\cite{Iacocca2016,Iacocca2017c,Iacocca2020} and significant developments have been achieved by measuring their ferromagnetic resonance~\cite{Jungfleisch2016,Bhat2016,Negrello2022}, including patterning modifications to identify magnetization states~\cite{Arroo2019,Dion2019,Gartside2021,Goryca2021,Vanstone2022} as well as reservoir computing applications~\cite{Gartside2022,Stenning2024}. More recently, it has been shown that ASIs can be driven to nonlinear dynamics~\cite{Lendinez2023} and can be realized in three-dimensional (3D) geometries~\cite{May2021,Sahoo2021,Dion2024,Berchialla2024}.

Much of the thermodynamical properties of these geometrical arrangements can be realized on macroscopic scales~\cite{Olive1998}. In fact, periodic lattices made of bar magnets have shown frustration and similar ground states~\cite{Mellado2012,Teixeira2024}. Recently, a macroscopic artificial spin ice, or macro-ASI, was realized to investigate resonant dynamics of mechano-magnetic systems~\cite{Peroor2024}. The combined experimental and numerical study demonstrated that the coupling between magnetic dipolar and mechanical degrees of freedom in a macro-ASI leads to rich nonlinear dynamics, including the stabilization of a frequency comb. Such findings could potentially be used to design and improve the coupling and nonlinear features of micro-mechanical resonators.

The macro-ASI provides an excellent testbed for investigating complex dynamics, including frustration. In particular, if any application is to be derived from ASIs, it would be advantageous to develop a mechanism to control wave transport through them. In electrical systems, transistors provide the machinery to manage the flow of electrical current. In magnonic systems, a promising solution could instead be topological defects embedded in ASIs. Dirac strings, in particular, could condition wave dynamics in ASIs by making certain paths through the lattice more favorable to propagation than others. This approach is reminiscent of one of the first investigations of magnetization dynamics in nanoscopic ASIs, where topological defects were found to give rise to new modes in the ferromagnetic spectrum associated with GHz dynamics localized at the edges of the individual nanomagnets~\cite{Gliga2013}. In a macro-ASI, the permanent magnets preclude localized dynamical modes and instead each bar magnet can be considered to be an ideal dipole to a good approximation. However, Dirac strings can also be stabilized within the lattice, resulting in a spatial modification of the propagating waves.

Here, we present numerical simulations of wave dynamics in a macro-ASI and investigate the impact of a Dirac string on its characteristics. We first determine that the macro-ASI supports two bands in its first Brillouin zone that are degenerate when propagating along the $x$ and $y$ directions. These bands define the range in which standing waves are stabilized in a physically constrained macro-ASI. The strong coupling between magnets leads to significant scattering between such modes, and excited waves quickly lose coherence along the lattice. The inclusion of a Dirac string does not arrest this behavior, but instead wave scattering excites localized modes within the Dirac string. This result demonstrates the direct correspondence between the dynamics of mechanical artificial spin ices and their nanoscopic counterparts.

\section{Simulation geometry}

\begin{figure}[t]
\centering
\includegraphics[width=3.3in]{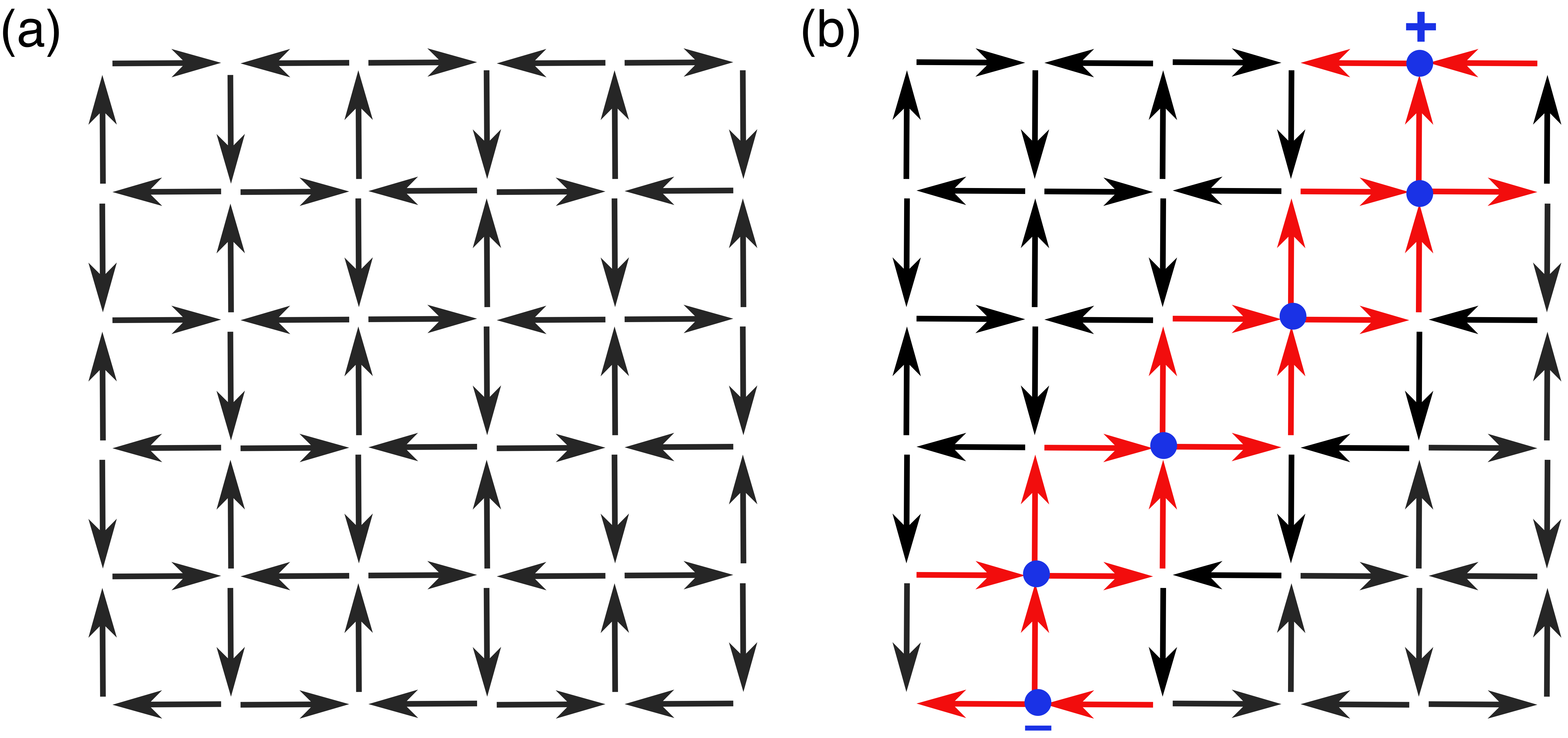}
\caption{\label{fig:Dirac_string_schematic} Schematic of the macro-ASI, with arrows pointing from each magnet's south pole to its north pole. (a) Represents the lattice in its ground state. (b) Represents the defective lattice, with blue dots indicating emergent monopoles tracing a Dirac string. The magnets highlighted in red indicate type-II couplings at the Dirac string.}
\end{figure}

We consider an array of sixty 2.54-cm-long magnets arranged in a square geometry with a lattice constant of $d=5.08$~cm. This arrangement is based on the physical macro-ASI built in Ref.~\cite{Peroor2024} where each magnet is supported by a low-friction rotary mount. At each vertex, four magnets strongly interact with each other. For our finite square arrangement, a global energy minimum is achieved in the system, corresponding to the type-I or vortex configuration~\cite{Wang2006}. This configuration is shown in Fig.~\ref{fig:Dirac_string_schematic}(a). A Dirac string can be set in the system by enforcing two vortex configurations with a type-II boundary. This is shown in Fig.~\ref{fig:Dirac_string_schematic}(b). The black arrows represent the magnets in the vortex state and the red arrows represent the flipped magnets. The energy in these magnets is higher since both horizontal and vertical magnets have the same orientation. This chain of ``reversed'' magnets defines higher energy vertex illustrated in blue dots that compose the Dirac string. The Dirac string makes it possible to connect the positive and negative emergent monopoles located at the physical edges of the macro-ASI.

The macro-ASI is numerically modeled using the magnetic-monopole approximation detailed in Refs.~\cite{Mellado2012,Teixeira2024,Peroor2024}, according to which each magnet is reduced to two monopoles carrying magnetic charges, \(+q\) and \(-q\). All magnets are governed by a Coulomb-like interaction among monopoles and energy dissipation due to friction at the rotors. Additionally, the magnets may be driven by an external magnetic flux density $B$. The resulting torque experienced by a magnet \(i\) is given by:
\begin{eqnarray}
\label{eq:SItorque}
    I\frac{d^2\theta_i}{dt^2}&=&-\eta\frac{d\theta_i}{dt}+\Bigg[q\textbf{L}_{\hat{\mu}_i}\times\textbf{B}(x,y,z)\\
    &&+\sum_{j=1,j\neq i}^N{\frac{\mu_0}{4\pi}\frac{q^2}{|\textbf{r}_{\hat{\mu}_i,\hat{\mu}_j}|^3}(\textbf{L}_{\hat{\mu}_i}\times\textbf{r}_{\hat{\mu}_i,\hat{\mu}_j})}\Bigg]\cdot\left(\hat{\mu}_i\times\hat{k}\right)\nonumber
\end{eqnarray}
where $I=2.03\times10^{-8}$~kg~m$^2$ is the moment of inertia of the bar magnet, $\eta=10^{-7}$ is the friction coefficient, and $q=2.08$~A~m is the effective monopole charge corresponding to the nominal saturation magnetization of the bar magnets, $M_s=1050$~kA/m. The third and fourth terms in Eq.~\eqref{eq:SItorque} are in general defined in three dimensions but the effective torque is constrained to a rotation axis by the rotary mount. To numerically represent this situation, we define the magnet's plane of rotation as $\hat{\mu}_i\times\hat{k}$, the orientation of the magnet as $\mathbf{L}_{\hat{\mu}_i}$, and the distance between monopoles by $\textbf{r}_{\hat{\mu}_i,\hat{\mu}_j}=\mathbf{L}_{\hat{\mu}_j}+\mathbf{d}_l-\mathbf{L}_{\hat{\mu}_i}$, where $\mathbf{d}_l$ is the center-to-center distance between magnets $i$ and $j$.

\begin{figure}[t]
\centering
\includegraphics[width=3.3in]{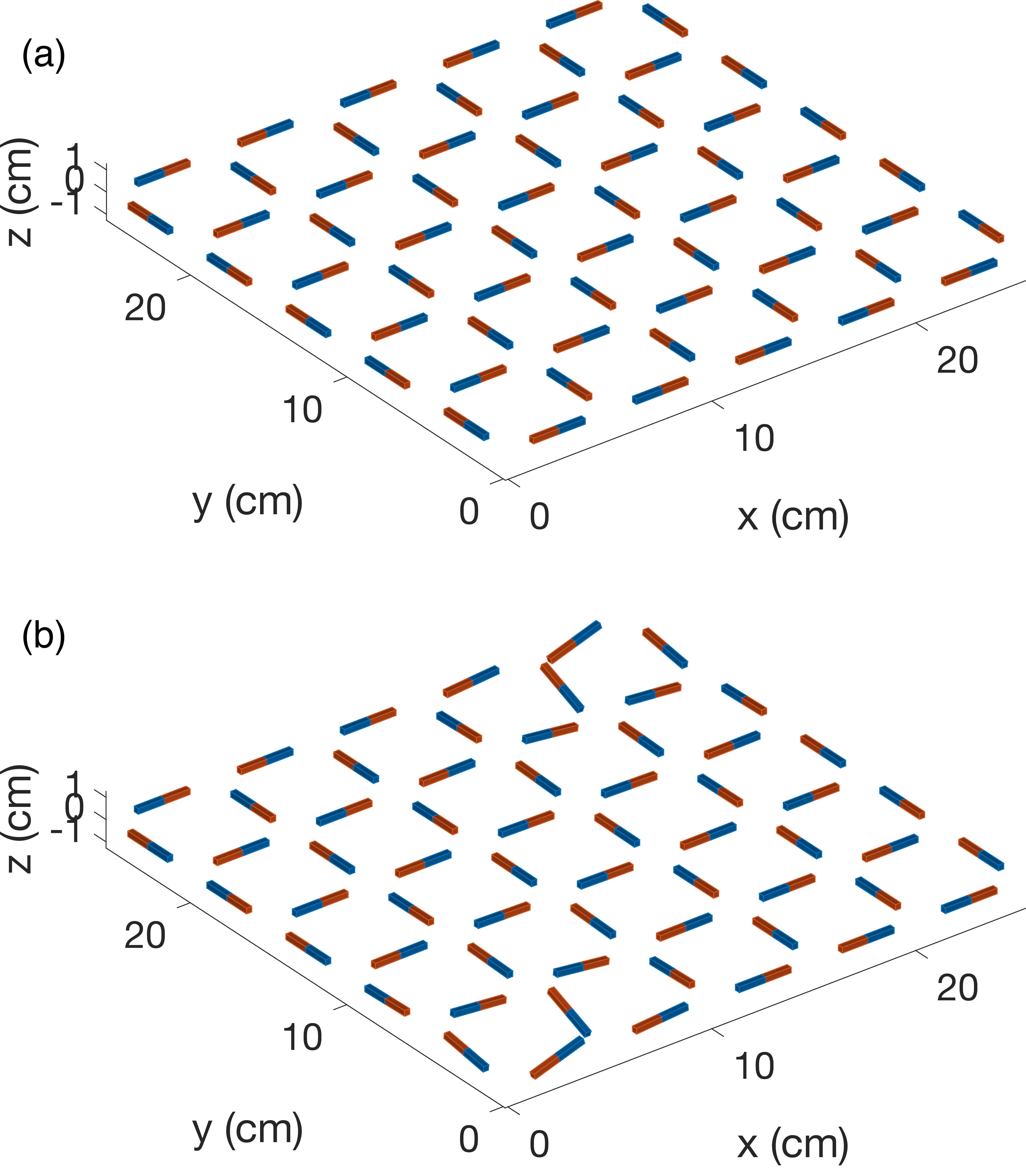}
\caption{\label{fig:Linked_plots} Stabilized ASI states for the (a) ground state and (b) defective state with a Dirac string across the macro-ASI's diagonal.}
\end{figure}

Equation~\eqref{eq:SItorque} was solved using MATLAB's ode15s, applying a variable-step variable-order method to a system of stiff differential equations. To stabilize an appropriate ground state, we provided the configurations shown in Fig.~\ref{fig:Dirac_string_schematic} with an additional small random noise as initial conditions. This state was allowed to relax for 20 seconds. Then, a small 10~mT spatially-uniform magnetic field was applied for 10 seconds and subsequently removed for another 10 seconds. Following this three-step procedure provided a stable initial state for both the ground state and defective states for dynamic simulations. The relaxed states are shown in Fig.~\ref{fig:Linked_plots}. Note that there is little vertical deviation for the magnets along the Dirac string except for the physical edges where the emergent monopoles are located, c.f. Fig.~\ref{fig:Dirac_string_schematic}(b).

\section{Mechano-magnetic waves dispersion}

While the system we aim to explore is confined, the dispersion relation provides insights into the wave anisotropy and band structure in the macro-ASI model. For this, we perform numerical simulations in an extended macro-ASI of $25\times25$ lattice sites, i.e., 650 magnets, relaxed to their ground state configuration. This extended array provides sufficient wavevector resolution in reciprocal space.

Waves are excited by rotating one magnet close to the center of the array by 20 degrees. This is equivalent to a delta-function excitation which develops into every possible wavelength in the system. Therefore, Fourier transformation of the spatial and temporal data results in the system's dispersion relation~\cite{Venkat2013}. For this, we record the time evolution of the angle $\theta_i(t)$ for each magnet in time and then map the angles onto a regular grid, so that we can define $\theta(x,y,t)$. Because of the discrete locations of the magnets, this matrix exhibits a ``checkerboard'' pattern where data is present. This implies that the Fourier transform of such an array will be difficult to compute accurately stemming from the convolution theorem. In other words, the checkerboard pattern can be assumed to be a mask that is convolved with the Fourier transform of the ``true'' data. To solve this issue, we interpolate $\theta(x,y,t)$ in space at each time step. The best results were obtained with MATLAB's ``natural'' 2D interpolation. To avoid numerically induced anisotropy in the reconstruction, we computed the Fourier transform for datasets rotated by 90~degrees. From this treatment, we obtain $\hat{\theta}(k_x,k_y,\omega)$.

\begin{figure}[t]
\centering
\includegraphics[width=3.3in]{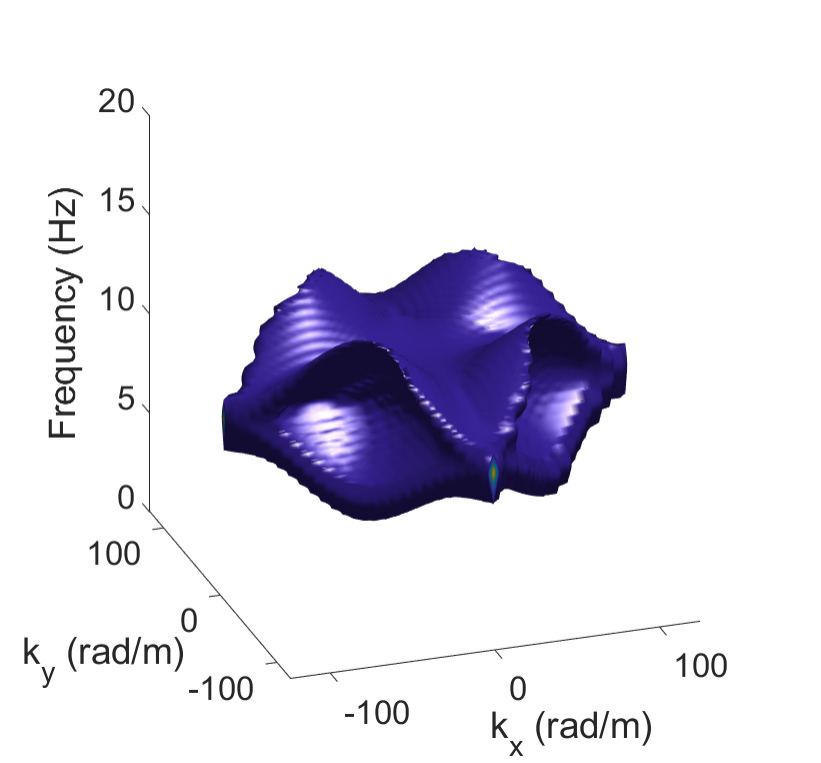}
\caption{\label{fig:dispersion} Dispersion relation for the macro-ASI. The contour exhibits two bands with four-fold symmetry except for the degenerate directions when $k_x=0$ or $k_y=0$.}
\end{figure}

For the ground state configuration, the primitive cell is composed of four nanomagnets interacting at one vertex. The lattice is then built from the primitive cell by the translation vectors $\vec{a}_1=2d\hat{x}$ and $\vec{a}_2=d(\hat{x}+\hat{y})$. This implies that the first Brillouin zone (FBZ) is given by the reciprocal vectors $\vec{b}_1=(\pi/d)(\hat{k}_x+\hat{k}_y)$ and $\vec{b}_2=(2\pi/d)\hat{k}_x$. The resulting dispersion relation in the FBZ is shown in Fig.~\ref{fig:dispersion} as a contour of the 3D Fourier transform data.
\begin{figure}[t]
\centering
\includegraphics[width=3.in]{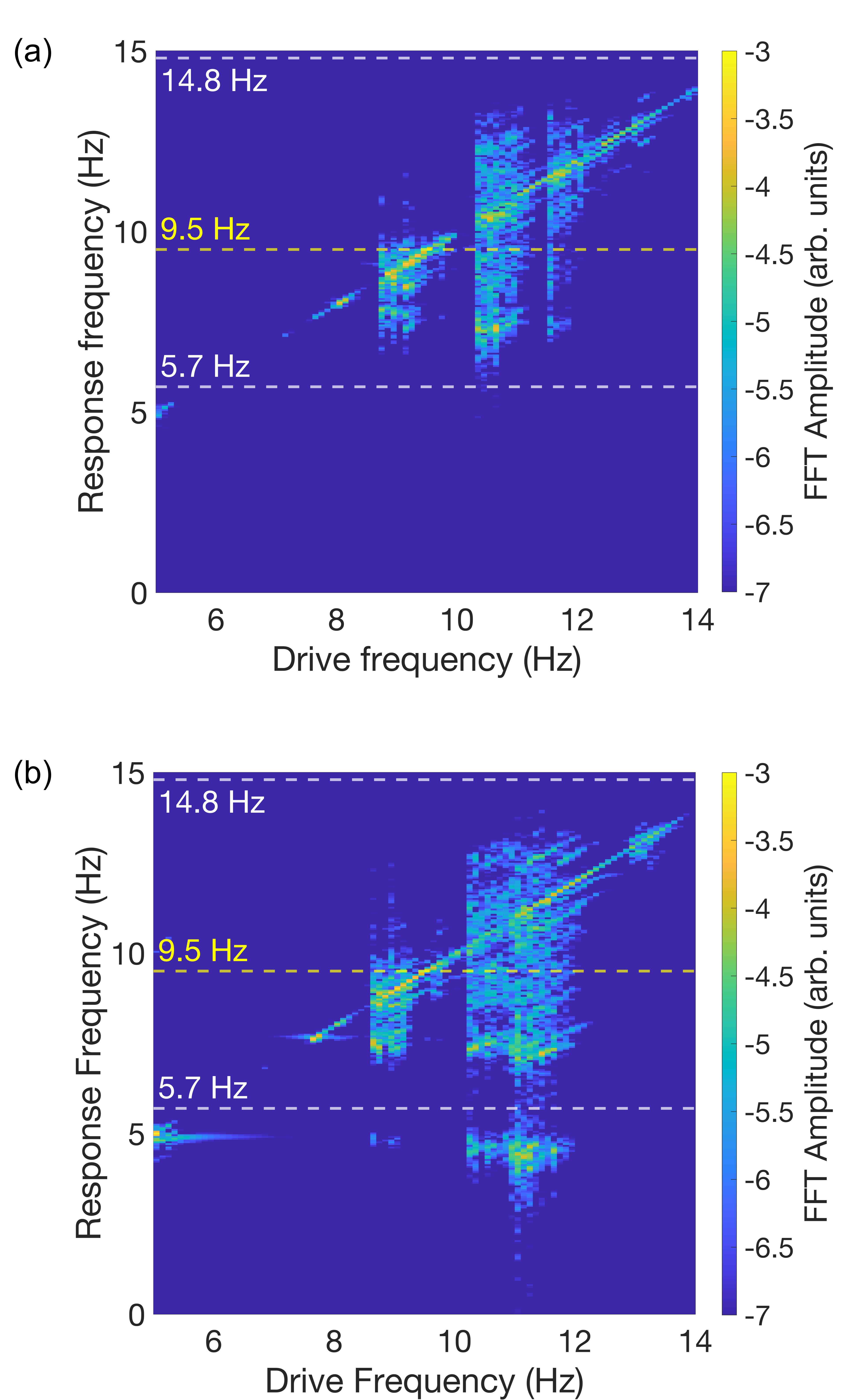}
\caption{Frequency spectra of one output magnet for varying excitation frequencies of the field applied to the input magnets shown for the (a) ground state and (b) defective state.}
\label{fig:Signal_Purity}
\end{figure}

It is immediately apparent that the macro-ASI has four-fold symmetry in the ground state, as expected. The bands are defined within $\approx5.7$~Hz and $\approx14.8$~Hz. The waves are degenerate and flat along $k_x=0$ and $k_y=0$, with a frequency of $\approx9.5$~Hz. This means that wave propagation in the macro-ASI is preferentially along diagonals. This intuition is in qualitative agreement with the standing wave modes observed in the confined macro-ASI~\cite{Peroor2024}.

\section{Forced dynamics}

In this section, we investigate wave propagation in the confined macro-ASI and the influence of a Dirac string. Due to the preferential diagonal wave propagation, we excite two magnets in one corner of the macro-ASI by an external oscillatory magnetic field, which we refer to as the input. The remainder of the magnets are not affected by the field and interact only through non-local dipolar forces. The output signal is collected from the magnets opposite to the input. Because of the four-fold symmetry of the system, any pair of diagonal magnets leads to the same dynamics.

First, we perform simulations by setting the amplitude of the oscillatory magnetic field to $B=1$~mT and vary its frequency from $5$~Hz to $14$~Hz in increments of $0.1$~Hz. In these simulations, we probed the output every $0.02$~s and ran the simulation for $20$~s, resulting in a frequency resolution of $50$~mHz and Nyquist frequency of $25$~Hz. The resulting spectra for the output magnet are shown in Fig.~\ref{fig:Signal_Purity}(a). While there is some discernible output at 5~Hz, a clear output is observed above $5.7$~Hz, over the minimum wave propagation frequency determined in Section II. The spectrum at the output becomes particularly broad past 10~Hz which is an indication of wave scattering in the lattice as well as formation of standing wave patterns. Notice that the broadening is contained within the band of propagating waves, as is expected for the formation of and scattering between localized modes. This is in agreement with features observed in Ref.~\cite{Peroor2024} although at different frequencies because of the choice of $M_s$. Despite this broadening, and thus loss of coherence, the output is dominated by the driving frequency.

We repeat the simulation when a Dirac string is stabilized, as shown in Fig.~\ref{fig:Linked_plots}(b). The excitation is in the corner normal to the Dirac string so we expect the wave to preferentially collide with the Dirac string. The resulting output spectra are shown in Fig.~\ref{fig:Signal_Purity}(b). The features are rather similar to Fig.~\ref{fig:Signal_Purity}(a), with a dominant linear output and loss of coherence past $10$~Hz. However, there is a clear flat signal at $\approx4.9$~Hz. This frequency is under the allowed wave frequencies and thus we surmise that it corresponds to localized modes. The natural question is to determine whether this localization occurs at the edges of the array or at the Dirac string.
\begin{figure}[t]
\centering
\includegraphics[width=3.3in]{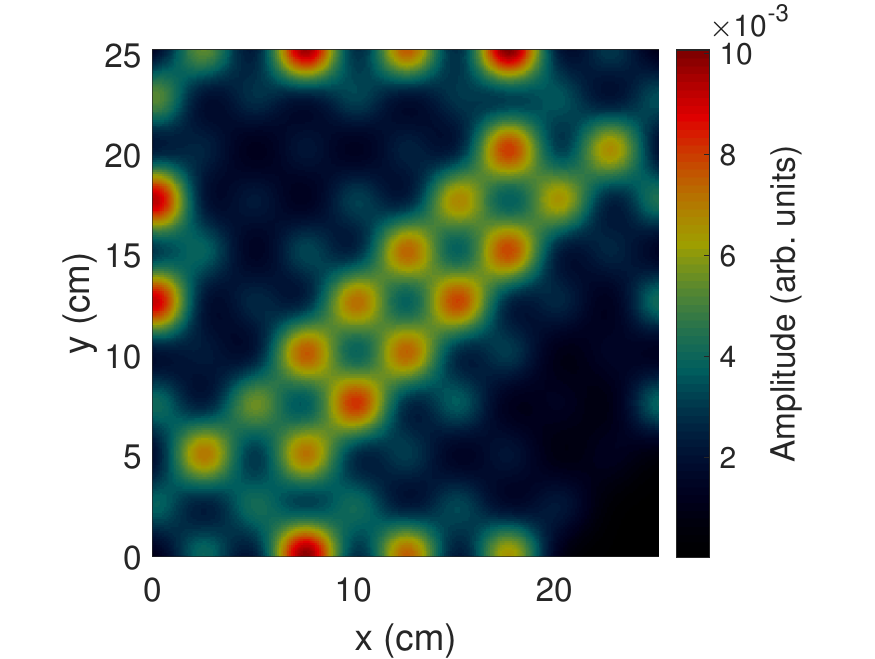}
\caption{The Dirac-string mode profile as an average of modes amplitudes between 4 Hz and 6 Hz.}
\label{fig:modes}
\end{figure}

To determine the spatial location of the mode, we perform a dedicated simulation at $11.2$~Hz and compute the Fourier transform for the time trace of each magnet in the array. We then determine the Fourier amplitude for each magnet and convolve it with a Gaussian function~\cite{Peroor2024}. This allows us to clearly visualize the mode volume. The mode profile at frequencies under $5.7$~Hz is shown in Fig.\ref{fig:modes}. Because of its broad spectrum, we averaged the mode amplitudes in the frequency range from 4~Hz to 6~Hz. Clearly, the mode is located at the Dirac string and is thus related to the different energy landscape due to the type-II vertices. Some evidence of localization at the physical edges is also observed. Because our system has no mechanism to impart chirality, all edges are excited.

\begin{figure}[t]
\centering
\includegraphics[width=3.3in]{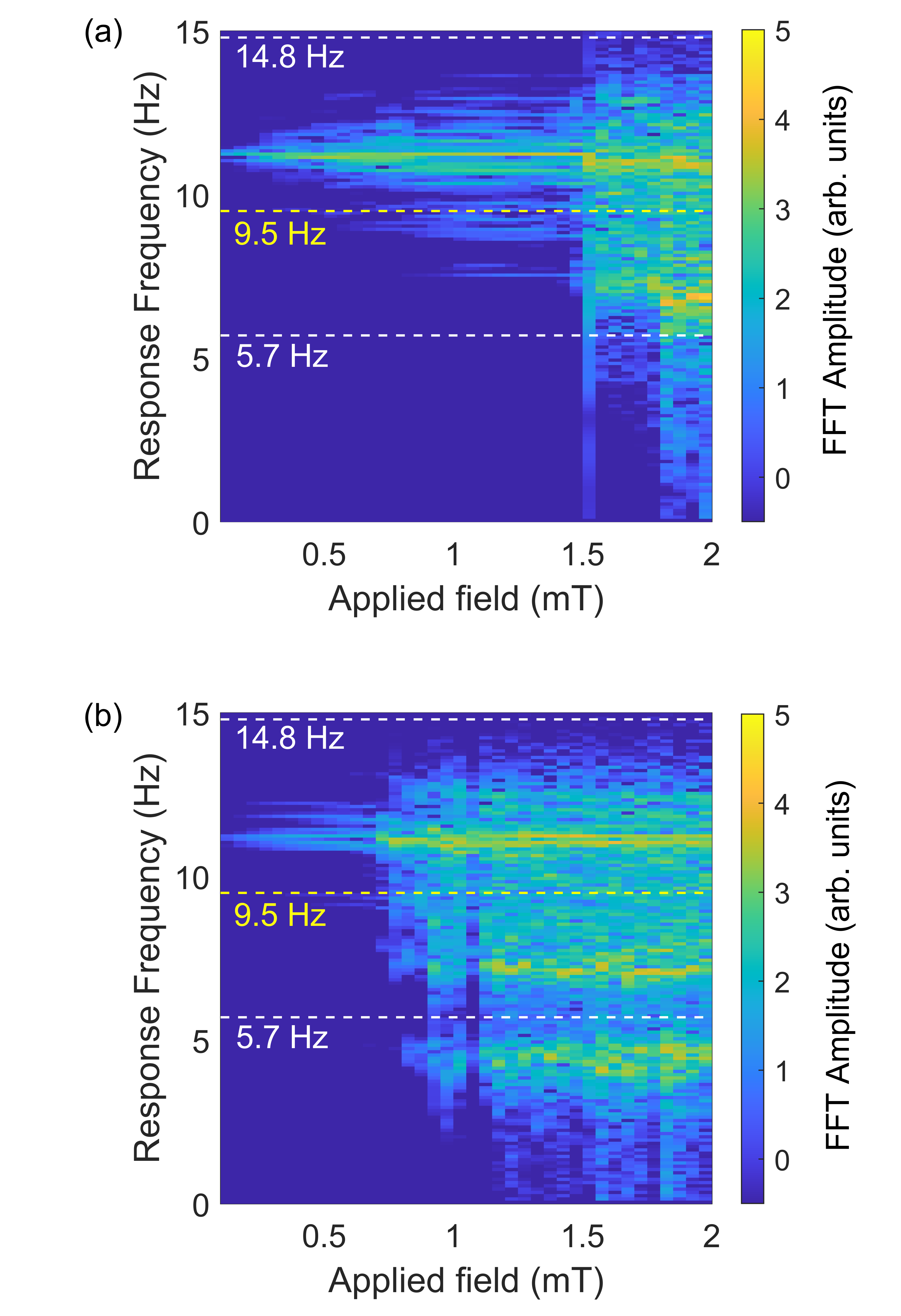}
\caption{Frequency spectra of one output magnet as a function of applied field magnitude and a frequency of $11.2$~Hz shown for the (a) ground state and (b) defective state.}
\label{fig:bdep}
\end{figure}
For the mechano-magnetic wave scattering to occur between the wave bath and the localized Dirac string, we expect nonlinearities to play an important role. To determine the onset of nonlinearities, we perform simulations as a function of applied field magnitude and at a frequency of $11.2$~Hz, where the scattering to the localized Dirac mode is observed. The results are shown in Fig.~\ref{fig:bdep} for the (a) ground state and (b) defective state. From Fig.~\ref{fig:bdep}(b), we see that the onset of the localized mode occurs at approximately $B=0.8$~mT. This is accompanied by a significant spectral broadening between $5.7$~Hz and $9.5$~Hz, the lower wave mode band. In contrast, the ground state only exhibits a modest spectral peak at $7$~Hz for comparable field magnitudes while a broad spectrum is observed starting at $B=1.8$~mT.

\section{Conclusion}

We have investigated the impact of a Dirac string on the dynamics of a macro-ASI. We find that this system can excite two wave bands that become degenerate in the $x$ and $y$ direction of propagation. This band structure allows for the formation of standing wave patterns and also inter-modal scattering, resulting in broad spectra. When a Dirac string is stabilized within the macro-ASI, wave scattering stimulates resonant dynamics in the Dirac string. Because of its spatial localization, such a resonant mode has a characteristic frequency below the wave dispersion bands. These results are in analogy with the appearance of low-frequency features in the FMR spectra of nanoscopic artificial spin ices due to emergent monopoles~\cite{Gliga2013}. However, our macro-ASI does not exhibit edge modes because the magnets are effectively permanent. The appearance of localized modes is thus related to the different coupling strengths between type-II and type-I vertices.

Contrary to expectations, the excited wave does not propagate towards the Dirac string and scatter on it. The excited wave instead scatters throughout the lattice, exciting local resonating modes at the Dirac string. This implies that Dirac strings cannot be directly used as simple ``gates'' for mechano-magnetic waves in artificial spin ices. The mode localization, however, can potentially channel excitations through the Dirac string. This would be similar to the spin wave propagation where domain walls act as nanoscale waveguides~\cite{Wagner2016}. However, it is important to recognize that the macro-ASI displays a significant loss in coherence in this process. Therefore, an excitation directly at the Dirac string resonant frequency may be a pathway to channel waves with better coherence.

\section*{Acknowledgments}

This material is based upon work supported by the National Science Foundation under Grant No. 2205796 (L.A.S. and E.I.) and Grant No. 2338060 (D.A.B.).


\begin{thebibliography}{31}%
\makeatletter
\providecommand \@ifxundefined [1]{%
 \@ifx{#1\undefined}
}%
\providecommand \@ifnum [1]{%
 \ifnum #1\expandafter \@firstoftwo
 \else \expandafter \@secondoftwo
 \fi
}%
\providecommand \@ifx [1]{%
 \ifx #1\expandafter \@firstoftwo
 \else \expandafter \@secondoftwo
 \fi
}%
\providecommand \natexlab [1]{#1}%
\providecommand \enquote  [1]{``#1''}%
\providecommand \bibnamefont  [1]{#1}%
\providecommand \bibfnamefont [1]{#1}%
\providecommand \citenamefont [1]{#1}%
\providecommand \href@noop [0]{\@secondoftwo}%
\providecommand \href [0]{\begingroup \@sanitize@url \@href}%
\providecommand \@href[1]{\@@startlink{#1}\@@href}%
\providecommand \@@href[1]{\endgroup#1\@@endlink}%
\providecommand \@sanitize@url [0]{\catcode `\\12\catcode `\$12\catcode
  `\&12\catcode `\#12\catcode `\^12\catcode `\_12\catcode `\%12\relax}%
\providecommand \@@startlink[1]{}%
\providecommand \@@endlink[0]{}%
\providecommand \url  [0]{\begingroup\@sanitize@url \@url }%
\providecommand \@url [1]{\endgroup\@href {#1}{\urlprefix }}%
\providecommand \urlprefix  [0]{URL }%
\providecommand \Eprint [0]{\href }%
\providecommand \doibase [0]{https://doi.org/}%
\providecommand \selectlanguage [0]{\@gobble}%
\providecommand \bibinfo  [0]{\@secondoftwo}%
\providecommand \bibfield  [0]{\@secondoftwo}%
\providecommand \translation [1]{[#1]}%
\providecommand \BibitemOpen [0]{}%
\providecommand \bibitemStop [0]{}%
\providecommand \bibitemNoStop [0]{.\EOS\space}%
\providecommand \EOS [0]{\spacefactor3000\relax}%
\providecommand \BibitemShut  [1]{\csname bibitem#1\endcsname}%
\let\auto@bib@innerbib\@empty
\bibitem [{\citenamefont {Skj{\ae}rv{\o}}\ \emph {et~al.}(2020)\citenamefont
  {Skj{\ae}rv{\o}}, \citenamefont {Marrows}, \citenamefont {Stamps},\ and\
  \citenamefont {Heyderman}}]{Skjaervo2020}%
  \BibitemOpen
  \bibfield  {author} {\bibinfo {author} {\bibfnamefont {S.~H.}\ \bibnamefont
  {Skj{\ae}rv{\o}}}, \bibinfo {author} {\bibfnamefont {C.~H.}\ \bibnamefont
  {Marrows}}, \bibinfo {author} {\bibfnamefont {R.~L.}\ \bibnamefont
  {Stamps}},\ and\ \bibinfo {author} {\bibfnamefont {L.~J.}\ \bibnamefont
  {Heyderman}},\ }\bibfield  {title} {\bibinfo {title} {Advances in artificial
  spin ice},\ }\href {https://www.nature.com/articles/s42254-019-0118-3}
  {\bibfield  {journal} {\bibinfo  {journal} {Nat. Rev. Phys.}\ }\textbf
  {\bibinfo {volume} {2}},\ \bibinfo {pages} {13} (\bibinfo {year}
  {2020})}\BibitemShut {NoStop}%
\bibitem [{\citenamefont {Heyderman}\ and\ \citenamefont
  {Stamps}(2013)}]{Heyderman2013}%
  \BibitemOpen
  \bibfield  {author} {\bibinfo {author} {\bibfnamefont {L.~J.}\ \bibnamefont
  {Heyderman}}\ and\ \bibinfo {author} {\bibfnamefont {R.~L.}\ \bibnamefont
  {Stamps}},\ }\bibfield  {title} {\bibinfo {title} {Artificial ferroic
  systems: novel functionality from structure, interactions and dynamics},\
  }\href {http://stacks.iop.org/0953-8984/25/i=36/a=363201} {\bibfield
  {journal} {\bibinfo  {journal} {Journal of Physics: Condensed Matter}\
  }\textbf {\bibinfo {volume} {25}},\ \bibinfo {pages} {363201} (\bibinfo
  {year} {2013})}\BibitemShut {NoStop}%
\bibitem [{\citenamefont {Wang}\ \emph {et~al.}(2006)\citenamefont {Wang},
  \citenamefont {Nisoli}, \citenamefont {Freitas}, \citenamefont {Li},
  \citenamefont {McConville}, \citenamefont {Cooley}, \citenamefont {Lund},
  \citenamefont {Samarth}, \citenamefont {Leighton}, \citenamefont {Crespi},\
  and\ \citenamefont {Schiffer}}]{Wang2006}%
  \BibitemOpen
  \bibfield  {author} {\bibinfo {author} {\bibfnamefont {R.}~\bibnamefont
  {Wang}}, \bibinfo {author} {\bibfnamefont {C.}~\bibnamefont {Nisoli}},
  \bibinfo {author} {\bibfnamefont {R.}~\bibnamefont {Freitas}}, \bibinfo
  {author} {\bibfnamefont {J.}~\bibnamefont {Li}}, \bibinfo {author}
  {\bibfnamefont {W.}~\bibnamefont {McConville}}, \bibinfo {author}
  {\bibfnamefont {B.}~\bibnamefont {Cooley}}, \bibinfo {author} {\bibfnamefont
  {M.}~\bibnamefont {Lund}}, \bibinfo {author} {\bibfnamefont {N.}~\bibnamefont
  {Samarth}}, \bibinfo {author} {\bibfnamefont {C.}~\bibnamefont {Leighton}},
  \bibinfo {author} {\bibfnamefont {V.}~\bibnamefont {Crespi}},\ and\ \bibinfo
  {author} {\bibfnamefont {P.}~\bibnamefont {Schiffer}},\ }\bibfield  {title}
  {\bibinfo {title} {Artificial spin ice in a geometrical frustrated lattice of
  nanoscale ferromagnetic islands},\ }\href
  {http://www.nature.com/nature/journal/v439/n7074/full/nature04447.html}
  {\bibfield  {journal} {\bibinfo  {journal} {Nature}\ }\textbf {\bibinfo
  {volume} {439}},\ \bibinfo {pages} {303} (\bibinfo {year}
  {2006})}\BibitemShut {NoStop}%
\bibitem [{\citenamefont {Gilbert}\ \emph {et~al.}(2014)\citenamefont
  {Gilbert}, \citenamefont {Chern}, \citenamefont {Zhange}, \citenamefont
  {O'Brien}, \citenamefont {Foe}, \citenamefont {Nisoli},\ and\ \citenamefont
  {Schiffer}}]{Gilbert2014}%
  \BibitemOpen
  \bibfield  {author} {\bibinfo {author} {\bibfnamefont {I.}~\bibnamefont
  {Gilbert}}, \bibinfo {author} {\bibfnamefont {G.-W.}\ \bibnamefont {Chern}},
  \bibinfo {author} {\bibfnamefont {S.}~\bibnamefont {Zhange}}, \bibinfo
  {author} {\bibfnamefont {L.}~\bibnamefont {O'Brien}}, \bibinfo {author}
  {\bibfnamefont {B.}~\bibnamefont {Foe}}, \bibinfo {author} {\bibfnamefont
  {C.}~\bibnamefont {Nisoli}},\ and\ \bibinfo {author} {\bibfnamefont
  {P.}~\bibnamefont {Schiffer}},\ }\bibfield  {title} {\bibinfo {title}
  {Emergent ice rule and magnetic charge screening from vertex frustration in
  artificial spin ice},\ }\href {https://doi.org/10.1038/nphys3037} {\bibfield
  {journal} {\bibinfo  {journal} {Nature Physics}\ }\textbf {\bibinfo {volume}
  {10}},\ \bibinfo {pages} {670} (\bibinfo {year} {2014})}\BibitemShut
  {NoStop}%
\bibitem [{\citenamefont {Gliga}\ \emph {et~al.}(2020)\citenamefont {Gliga},
  \citenamefont {Iacocca},\ and\ \citenamefont {Heinonen}}]{Gliga2020}%
  \BibitemOpen
  \bibfield  {author} {\bibinfo {author} {\bibfnamefont {S.}~\bibnamefont
  {Gliga}}, \bibinfo {author} {\bibfnamefont {E.}~\bibnamefont {Iacocca}},\
  and\ \bibinfo {author} {\bibfnamefont {O.~G.}\ \bibnamefont {Heinonen}},\
  }\bibfield  {title} {\bibinfo {title} {Dynamics of reconfigurable artificial
  spin ice: Toward magnonic functional materials},\ }\href
  {https://doi.org/10.1063/1.5142705} {\bibfield  {journal} {\bibinfo
  {journal} {APL Materials}\ }\textbf {\bibinfo {volume} {8}},\ \bibinfo
  {pages} {040911} (\bibinfo {year} {2020})},\ \Eprint
  {https://arxiv.org/abs/https://doi.org/10.1063/1.5142705}
  {https://doi.org/10.1063/1.5142705} \BibitemShut {NoStop}%
\bibitem [{\citenamefont {Lendinez}\ and\ \citenamefont
  {Jungfleisch}(2019)}]{Lendinez2019}%
  \BibitemOpen
  \bibfield  {author} {\bibinfo {author} {\bibfnamefont {S.}~\bibnamefont
  {Lendinez}}\ and\ \bibinfo {author} {\bibfnamefont {M.~B.}\ \bibnamefont
  {Jungfleisch}},\ }\bibfield  {title} {\bibinfo {title} {Magnetization
  dynamics in artificial spin ice},\ }\href
  {https://iopscience.iop.org/article/10.1088/1361-648X/ab3e78} {\bibfield
  {journal} {\bibinfo  {journal} {J. Phys.: Condens. Matter}\ }\textbf
  {\bibinfo {volume} {32}},\ \bibinfo {pages} {013001} (\bibinfo {year}
  {2019})}\BibitemShut {NoStop}%
\bibitem [{\citenamefont {Iacocca}\ \emph {et~al.}(2016)\citenamefont
  {Iacocca}, \citenamefont {Gliga}, \citenamefont {Stamps},\ and\ \citenamefont
  {Heinonen}}]{Iacocca2016}%
  \BibitemOpen
  \bibfield  {author} {\bibinfo {author} {\bibfnamefont {E.}~\bibnamefont
  {Iacocca}}, \bibinfo {author} {\bibfnamefont {S.}~\bibnamefont {Gliga}},
  \bibinfo {author} {\bibfnamefont {R.~L.}\ \bibnamefont {Stamps}},\ and\
  \bibinfo {author} {\bibfnamefont {O.}~\bibnamefont {Heinonen}},\ }\bibfield
  {title} {\bibinfo {title} {Reconfigurable wave band structure of an
  artificial square ice},\ }\href {https://doi.org/10.1103/PhysRevB.93.134420}
  {\bibfield  {journal} {\bibinfo  {journal} {Phys. Rev. B}\ }\textbf {\bibinfo
  {volume} {93}},\ \bibinfo {pages} {134420} (\bibinfo {year}
  {2016})}\BibitemShut {NoStop}%
\bibitem [{\citenamefont {Iacocca}\ and\ \citenamefont
  {Heinonen}(2017)}]{Iacocca2017c}%
  \BibitemOpen
  \bibfield  {author} {\bibinfo {author} {\bibfnamefont {E.}~\bibnamefont
  {Iacocca}}\ and\ \bibinfo {author} {\bibfnamefont {O.}~\bibnamefont
  {Heinonen}},\ }\bibfield  {title} {\bibinfo {title} {Topologically nontrivial
  magnon bands in artificial square spin ices with dzyaloshinskii-moriya
  interaction},\ }\href {https://doi.org/10.1103/PhysRevApplied.8.034015}
  {\bibfield  {journal} {\bibinfo  {journal} {Phys. Rev. Applied}\ }\textbf
  {\bibinfo {volume} {8}},\ \bibinfo {pages} {034015} (\bibinfo {year}
  {2017})}\BibitemShut {NoStop}%
\bibitem [{\citenamefont {Iacocca}\ \emph {et~al.}(2020)\citenamefont
  {Iacocca}, \citenamefont {Gliga},\ and\ \citenamefont
  {Heinonen}}]{Iacocca2020}%
  \BibitemOpen
  \bibfield  {author} {\bibinfo {author} {\bibfnamefont {E.}~\bibnamefont
  {Iacocca}}, \bibinfo {author} {\bibfnamefont {S.}~\bibnamefont {Gliga}},\
  and\ \bibinfo {author} {\bibfnamefont {O.~G.}\ \bibnamefont {Heinonen}},\
  }\bibfield  {title} {\bibinfo {title} {Tailoring spin-wave channels in a
  reconfigurable artificial spin ice},\ }\href
  {https://doi.org/10.1103/PhysRevApplied.13.044047} {\bibfield  {journal}
  {\bibinfo  {journal} {Phys. Rev. Applied}\ }\textbf {\bibinfo {volume}
  {13}},\ \bibinfo {pages} {044047} (\bibinfo {year} {2020})}\BibitemShut
  {NoStop}%
\bibitem [{\citenamefont {Jungfleisch}\ \emph {et~al.}(2016)\citenamefont
  {Jungfleisch}, \citenamefont {Zhang}, \citenamefont {Iacocca}, \citenamefont
  {Sklenar}, \citenamefont {Ding}, \citenamefont {Jiang}, \citenamefont
  {Zhang}, \citenamefont {Pearson}, \citenamefont {Novosad}, \citenamefont
  {Ketterson}, \citenamefont {Heinonen},\ and\ \citenamefont
  {Hoffmann}}]{Jungfleisch2016}%
  \BibitemOpen
  \bibfield  {author} {\bibinfo {author} {\bibfnamefont {M.~B.}\ \bibnamefont
  {Jungfleisch}}, \bibinfo {author} {\bibfnamefont {W.}~\bibnamefont {Zhang}},
  \bibinfo {author} {\bibfnamefont {E.}~\bibnamefont {Iacocca}}, \bibinfo
  {author} {\bibfnamefont {J.}~\bibnamefont {Sklenar}}, \bibinfo {author}
  {\bibfnamefont {J.}~\bibnamefont {Ding}}, \bibinfo {author} {\bibfnamefont
  {W.}~\bibnamefont {Jiang}}, \bibinfo {author} {\bibfnamefont
  {S.}~\bibnamefont {Zhang}}, \bibinfo {author} {\bibfnamefont {J.~E.}\
  \bibnamefont {Pearson}}, \bibinfo {author} {\bibfnamefont {V.}~\bibnamefont
  {Novosad}}, \bibinfo {author} {\bibfnamefont {J.~B.}\ \bibnamefont
  {Ketterson}}, \bibinfo {author} {\bibfnamefont {O.}~\bibnamefont
  {Heinonen}},\ and\ \bibinfo {author} {\bibfnamefont {A.}~\bibnamefont
  {Hoffmann}},\ }\bibfield  {title} {\bibinfo {title} {Dynamic response of an
  artificial square spin ice},\ }\href
  {https://doi.org/10.1103/PhysRevB.93.100401} {\bibfield  {journal} {\bibinfo
  {journal} {Phys. Rev. B}\ }\textbf {\bibinfo {volume} {93}},\ \bibinfo
  {pages} {100401} (\bibinfo {year} {2016})}\BibitemShut {NoStop}%
\bibitem [{\citenamefont {Bhat}\ \emph {et~al.}(2016)\citenamefont {Bhat},
  \citenamefont {Heimbach}, \citenamefont {Stasinopoulos},\ and\ \citenamefont
  {Grundler}}]{Bhat2016}%
  \BibitemOpen
  \bibfield  {author} {\bibinfo {author} {\bibfnamefont {V.~S.}\ \bibnamefont
  {Bhat}}, \bibinfo {author} {\bibfnamefont {F.}~\bibnamefont {Heimbach}},
  \bibinfo {author} {\bibfnamefont {I.}~\bibnamefont {Stasinopoulos}},\ and\
  \bibinfo {author} {\bibfnamefont {D.}~\bibnamefont {Grundler}},\ }\bibfield
  {title} {\bibinfo {title} {Magnetization dynamics of topological defects and
  the spin solid in a kagome artificial spin ice},\ }\href
  {https://doi.org/10.1103/PhysRevB.93.140401} {\bibfield  {journal} {\bibinfo
  {journal} {Phys. Rev. B}\ }\textbf {\bibinfo {volume} {93}},\ \bibinfo
  {pages} {140401} (\bibinfo {year} {2016})}\BibitemShut {NoStop}%
\bibitem [{\citenamefont {Negrello}\ \emph {et~al.}(2022)\citenamefont
  {Negrello}, \citenamefont {Montoncello}, \citenamefont {Kaffash},
  \citenamefont {Jungfleisch},\ and\ \citenamefont {Gubbiotti}}]{Negrello2022}%
  \BibitemOpen
  \bibfield  {author} {\bibinfo {author} {\bibfnamefont {R.}~\bibnamefont
  {Negrello}}, \bibinfo {author} {\bibfnamefont {F.}~\bibnamefont
  {Montoncello}}, \bibinfo {author} {\bibfnamefont {M.~T.}\ \bibnamefont
  {Kaffash}}, \bibinfo {author} {\bibfnamefont {M.~B.}\ \bibnamefont
  {Jungfleisch}},\ and\ \bibinfo {author} {\bibfnamefont {G.}~\bibnamefont
  {Gubbiotti}},\ }\bibfield  {title} {\bibinfo {title} {Dynamic coupling and
  spin-wave dispersions in a magnetic hybrid system made of an artificial
  spin-ice structure and an extended nife underlayer},\ }\href
  {https://doi.org/10.1063/5.0102571} {\bibfield  {journal} {\bibinfo
  {journal} {APL Materials}\ }\textbf {\bibinfo {volume} {10}},\ \bibinfo
  {pages} {091115} (\bibinfo {year} {2022})}\BibitemShut {NoStop}%
\bibitem [{\citenamefont {Arroo}\ \emph {et~al.}(2019)\citenamefont {Arroo},
  \citenamefont {Gartside},\ and\ \citenamefont {Branford}}]{Arroo2019}%
  \BibitemOpen
  \bibfield  {author} {\bibinfo {author} {\bibfnamefont {D.~M.}\ \bibnamefont
  {Arroo}}, \bibinfo {author} {\bibfnamefont {J.~C.}\ \bibnamefont
  {Gartside}},\ and\ \bibinfo {author} {\bibfnamefont {W.~R.}\ \bibnamefont
  {Branford}},\ }\bibfield  {title} {\bibinfo {title} {Sculpting the spin-wave
  response of artificial spin ice via microstate selection},\ }\href
  {https://doi.org/10.1103/PhysRevB.100.214425} {\bibfield  {journal} {\bibinfo
   {journal} {Phys. Rev. B}\ }\textbf {\bibinfo {volume} {100}},\ \bibinfo
  {pages} {214425} (\bibinfo {year} {2019})}\BibitemShut {NoStop}%
\bibitem [{\citenamefont {Dion}\ \emph {et~al.}(2019)\citenamefont {Dion},
  \citenamefont {Arroo}, \citenamefont {Yamanoi}, \citenamefont {Kimura},
  \citenamefont {Gartside}, \citenamefont {Cohen}, \citenamefont
  {Kurebayashi},\ and\ \citenamefont {Branford}}]{Dion2019}%
  \BibitemOpen
  \bibfield  {author} {\bibinfo {author} {\bibfnamefont {T.}~\bibnamefont
  {Dion}}, \bibinfo {author} {\bibfnamefont {D.~M.}\ \bibnamefont {Arroo}},
  \bibinfo {author} {\bibfnamefont {K.}~\bibnamefont {Yamanoi}}, \bibinfo
  {author} {\bibfnamefont {T.}~\bibnamefont {Kimura}}, \bibinfo {author}
  {\bibfnamefont {J.~C.}\ \bibnamefont {Gartside}}, \bibinfo {author}
  {\bibfnamefont {L.~F.}\ \bibnamefont {Cohen}}, \bibinfo {author}
  {\bibfnamefont {H.}~\bibnamefont {Kurebayashi}},\ and\ \bibinfo {author}
  {\bibfnamefont {W.~R.}\ \bibnamefont {Branford}},\ }\bibfield  {title}
  {\bibinfo {title} {Tunable magnetization dynamics in artificial spin ice via
  shape anisotropy modification},\ }\href
  {https://doi.org/10.1103/PhysRevB.100.054433} {\bibfield  {journal} {\bibinfo
   {journal} {Phys. Rev. B}\ }\textbf {\bibinfo {volume} {100}},\ \bibinfo
  {pages} {054433} (\bibinfo {year} {2019})}\BibitemShut {NoStop}%
\bibitem [{\citenamefont {Gartside}\ \emph {et~al.}(2021)\citenamefont
  {Gartside}, \citenamefont {Vanstone}, \citenamefont {Dion}, \citenamefont
  {Stenning}, \citenamefont {Arroo}, \citenamefont {Kurebayashi},\ and\
  \citenamefont {Branford}}]{Gartside2021}%
  \BibitemOpen
  \bibfield  {author} {\bibinfo {author} {\bibfnamefont {J.~C.}\ \bibnamefont
  {Gartside}}, \bibinfo {author} {\bibfnamefont {A.}~\bibnamefont {Vanstone}},
  \bibinfo {author} {\bibfnamefont {T.}~\bibnamefont {Dion}}, \bibinfo {author}
  {\bibfnamefont {K.~D.}\ \bibnamefont {Stenning}}, \bibinfo {author}
  {\bibfnamefont {D.~M.}\ \bibnamefont {Arroo}}, \bibinfo {author}
  {\bibfnamefont {H.}~\bibnamefont {Kurebayashi}},\ and\ \bibinfo {author}
  {\bibfnamefont {W.~R.}\ \bibnamefont {Branford}},\ }\bibfield  {title}
  {\bibinfo {title} {Reconfigurable magnonic mode-hybridisation and spectral
  control in a bicomponnt artificial spin ice},\ }\href
  {https://www.nature.com/articles/s41467-021-22723-x} {\bibfield  {journal}
  {\bibinfo  {journal} {Nature Communications}\ }\textbf {\bibinfo {volume}
  {12}},\ \bibinfo {pages} {2488} (\bibinfo {year} {2021})}\BibitemShut
  {NoStop}%
\bibitem [{\citenamefont {Goryca}\ \emph {et~al.}(2021)\citenamefont {Goryca},
  \citenamefont {Zhang}, \citenamefont {Li}, \citenamefont {Balk},
  \citenamefont {Watts}, \citenamefont {Leighton}, \citenamefont {Nisoli},
  \citenamefont {Schiffer},\ and\ \citenamefont {Crooker}}]{Goryca2021}%
  \BibitemOpen
  \bibfield  {author} {\bibinfo {author} {\bibfnamefont {M.}~\bibnamefont
  {Goryca}}, \bibinfo {author} {\bibfnamefont {X.}~\bibnamefont {Zhang}},
  \bibinfo {author} {\bibfnamefont {J.}~\bibnamefont {Li}}, \bibinfo {author}
  {\bibfnamefont {A.~L.}\ \bibnamefont {Balk}}, \bibinfo {author}
  {\bibfnamefont {J.~D.}\ \bibnamefont {Watts}}, \bibinfo {author}
  {\bibfnamefont {C.}~\bibnamefont {Leighton}}, \bibinfo {author}
  {\bibfnamefont {C.}~\bibnamefont {Nisoli}}, \bibinfo {author} {\bibfnamefont
  {P.}~\bibnamefont {Schiffer}},\ and\ \bibinfo {author} {\bibfnamefont
  {S.~A.}\ \bibnamefont {Crooker}},\ }\bibfield  {title} {\bibinfo {title}
  {Field-induced magnetic monopole plasma in artificial spin ice},\ }\href
  {https://doi.org/10.1103/PhysRevX.11.011042} {\bibfield  {journal} {\bibinfo
  {journal} {Phys. Rev. X}\ }\textbf {\bibinfo {volume} {11}},\ \bibinfo
  {pages} {011042} (\bibinfo {year} {2021})}\BibitemShut {NoStop}%
\bibitem [{\citenamefont {Vanstone}\ \emph {et~al.}(2022)\citenamefont
  {Vanstone}, \citenamefont {Gartside}, \citenamefont {Stenning}, \citenamefont
  {Dion}, \citenamefont {Arroo},\ and\ \citenamefont
  {Branford}}]{Vanstone2022}%
  \BibitemOpen
  \bibfield  {author} {\bibinfo {author} {\bibfnamefont {A.}~\bibnamefont
  {Vanstone}}, \bibinfo {author} {\bibfnamefont {J.~C.}\ \bibnamefont
  {Gartside}}, \bibinfo {author} {\bibfnamefont {K.~D.}\ \bibnamefont
  {Stenning}}, \bibinfo {author} {\bibfnamefont {T.}~\bibnamefont {Dion}},
  \bibinfo {author} {\bibfnamefont {D.~M.}\ \bibnamefont {Arroo}},\ and\
  \bibinfo {author} {\bibfnamefont {W.~R.}\ \bibnamefont {Branford}},\
  }\bibfield  {title} {\bibinfo {title} {Spectral fingerprinting: microstate
  readout via remanence ferromagnetic resonance in artificial spin ice},\
  }\href {https://doi.org/10.1088/1367-2630/ac608b} {\bibfield  {journal}
  {\bibinfo  {journal} {New Journal of Physics}\ }\textbf {\bibinfo {volume}
  {24}},\ \bibinfo {pages} {043017} (\bibinfo {year} {2022})}\BibitemShut
  {NoStop}%
\bibitem [{\citenamefont {Gartside}\ \emph {et~al.}(2022)\citenamefont
  {Gartside}, \citenamefont {Stenning}, \citenamefont {Vanstone}, \citenamefont
  {Holder}, \citenamefont {Arroo}, \citenamefont {Dion}, \citenamefont
  {Caravelli}, \citenamefont {Kurebayashi},\ and\ \citenamefont
  {Branford}}]{Gartside2022}%
  \BibitemOpen
  \bibfield  {author} {\bibinfo {author} {\bibfnamefont {J.~C.}\ \bibnamefont
  {Gartside}}, \bibinfo {author} {\bibfnamefont {K.~D.}\ \bibnamefont
  {Stenning}}, \bibinfo {author} {\bibfnamefont {A.}~\bibnamefont {Vanstone}},
  \bibinfo {author} {\bibfnamefont {H.~H.}\ \bibnamefont {Holder}}, \bibinfo
  {author} {\bibfnamefont {D.~M.}\ \bibnamefont {Arroo}}, \bibinfo {author}
  {\bibfnamefont {T.}~\bibnamefont {Dion}}, \bibinfo {author} {\bibfnamefont
  {F.}~\bibnamefont {Caravelli}}, \bibinfo {author} {\bibfnamefont
  {H.}~\bibnamefont {Kurebayashi}},\ and\ \bibinfo {author} {\bibfnamefont
  {W.~R.}\ \bibnamefont {Branford}},\ }\bibfield  {title} {\bibinfo {title}
  {Reconfigurable training and reservoir computing in an artificial spin-vortex
  ice via spin-wave fingerprinting},\ }\href
  {https://doi.org/10.1038/s41565-022-01091-7} {\bibfield  {journal} {\bibinfo
  {journal} {Nature Nanotechnology}\ }\textbf {\bibinfo {volume} {17}},\
  \bibinfo {pages} {460} (\bibinfo {year} {2022})}\BibitemShut {NoStop}%
\bibitem [{\citenamefont {Stenning}\ \emph {et~al.}(2024)\citenamefont
  {Stenning}, \citenamefont {Gartside}, \citenamefont {Manneschi},
  \citenamefont {Cheung}, \citenamefont {Chen}, \citenamefont {Vanstone},
  \citenamefont {Love}, \citenamefont {Holder}, \citenamefont {Caravelli},
  \citenamefont {Kurebayashi}, \citenamefont {Everschor-Sitte}, \citenamefont
  {Vasilaki},\ and\ \citenamefont {Branford}}]{Stenning2024}%
  \BibitemOpen
  \bibfield  {author} {\bibinfo {author} {\bibfnamefont {K.~D.}\ \bibnamefont
  {Stenning}}, \bibinfo {author} {\bibfnamefont {J.~C.}\ \bibnamefont
  {Gartside}}, \bibinfo {author} {\bibfnamefont {L.}~\bibnamefont {Manneschi}},
  \bibinfo {author} {\bibfnamefont {C.~T.~S.}\ \bibnamefont {Cheung}}, \bibinfo
  {author} {\bibfnamefont {T.}~\bibnamefont {Chen}}, \bibinfo {author}
  {\bibfnamefont {A.}~\bibnamefont {Vanstone}}, \bibinfo {author}
  {\bibfnamefont {J.}~\bibnamefont {Love}}, \bibinfo {author} {\bibfnamefont
  {H.}~\bibnamefont {Holder}}, \bibinfo {author} {\bibfnamefont
  {F.}~\bibnamefont {Caravelli}}, \bibinfo {author} {\bibfnamefont
  {H.}~\bibnamefont {Kurebayashi}}, \bibinfo {author} {\bibfnamefont
  {K.}~\bibnamefont {Everschor-Sitte}}, \bibinfo {author} {\bibfnamefont
  {E.}~\bibnamefont {Vasilaki}},\ and\ \bibinfo {author} {\bibfnamefont
  {W.~R.}\ \bibnamefont {Branford}},\ }\bibfield  {title} {\bibinfo {title}
  {Neuromorphic overparameterisation and few-shot learning in multilayer
  physical neural networks},\ }\href
  {https://doi.org/10.1038/s41467-024-50633-1} {\bibfield  {journal} {\bibinfo
  {journal} {Nature Communications}\ }\textbf {\bibinfo {volume} {15}},\
  \bibinfo {pages} {7377} (\bibinfo {year} {2024})}\BibitemShut {NoStop}%
\bibitem [{\citenamefont {Lendinez}\ \emph {et~al.}(2023)\citenamefont
  {Lendinez}, \citenamefont {Kaffash}, \citenamefont {Heinonen}, \citenamefont
  {Gliga}, \citenamefont {Iacocca},\ and\ \citenamefont
  {Jungfleisch}}]{Lendinez2023}%
  \BibitemOpen
  \bibfield  {author} {\bibinfo {author} {\bibfnamefont {S.}~\bibnamefont
  {Lendinez}}, \bibinfo {author} {\bibfnamefont {M.~T.}\ \bibnamefont
  {Kaffash}}, \bibinfo {author} {\bibfnamefont {O.~G.}\ \bibnamefont
  {Heinonen}}, \bibinfo {author} {\bibfnamefont {S.}~\bibnamefont {Gliga}},
  \bibinfo {author} {\bibfnamefont {E.}~\bibnamefont {Iacocca}},\ and\ \bibinfo
  {author} {\bibfnamefont {M.~B.}\ \bibnamefont {Jungfleisch}},\ }\bibfield
  {title} {\bibinfo {title} {Nonlinear multi-magnon scattering in artificial
  spin ice},\ }\href {https://www.nature.com/articles/s41467-023-38992-7}
  {\bibfield  {journal} {\bibinfo  {journal} {Nature Communications}\ }\textbf
  {\bibinfo {volume} {14}},\ \bibinfo {pages} {3419} (\bibinfo {year}
  {2023})}\BibitemShut {NoStop}%
\bibitem [{\citenamefont {May}\ \emph {et~al.}(2021)\citenamefont {May},
  \citenamefont {Saccone}, \citenamefont {van~den Berg}, \citenamefont {Askey},
  \citenamefont {Hunt},\ and\ \citenamefont {Ladak}}]{May2021}%
  \BibitemOpen
  \bibfield  {author} {\bibinfo {author} {\bibfnamefont {A.}~\bibnamefont
  {May}}, \bibinfo {author} {\bibfnamefont {M.}~\bibnamefont {Saccone}},
  \bibinfo {author} {\bibfnamefont {A.}~\bibnamefont {van~den Berg}}, \bibinfo
  {author} {\bibfnamefont {J.}~\bibnamefont {Askey}}, \bibinfo {author}
  {\bibfnamefont {M.}~\bibnamefont {Hunt}},\ and\ \bibinfo {author}
  {\bibfnamefont {S.}~\bibnamefont {Ladak}},\ }\bibfield  {title} {\bibinfo
  {title} {Magnetic charge propagation upon a 3d artificial spin ice},\ }\href
  {https://doi.org/10.1038/s41467-021-23480-7} {\bibfield  {journal} {\bibinfo
  {journal} {Nature Communications}\ }\textbf {\bibinfo {volume} {12}},\
  \bibinfo {pages} {3217} (\bibinfo {year} {2021})}\BibitemShut {NoStop}%
\bibitem [{\citenamefont {Sahoo}\ \emph {et~al.}(2021)\citenamefont {Sahoo},
  \citenamefont {May}, \citenamefont {van Den~Berg}, \citenamefont {Mondal},
  \citenamefont {Ladak},\ and\ \citenamefont {Barman}}]{Sahoo2021}%
  \BibitemOpen
  \bibfield  {author} {\bibinfo {author} {\bibfnamefont {S.}~\bibnamefont
  {Sahoo}}, \bibinfo {author} {\bibfnamefont {A.}~\bibnamefont {May}}, \bibinfo
  {author} {\bibfnamefont {A.}~\bibnamefont {van Den~Berg}}, \bibinfo {author}
  {\bibfnamefont {A.~K.}\ \bibnamefont {Mondal}}, \bibinfo {author}
  {\bibfnamefont {S.}~\bibnamefont {Ladak}},\ and\ \bibinfo {author}
  {\bibfnamefont {A.}~\bibnamefont {Barman}},\ }\bibfield  {title} {\bibinfo
  {title} {Observation of coherent spin waves in a three-dimensional artificial
  spin ice structure},\ }\href {https://doi.org/10.1021/acs.nanolett.1c00650}
  {\bibfield  {journal} {\bibinfo  {journal} {Nano Letters}\ }\textbf {\bibinfo
  {volume} {21}},\ \bibinfo {pages} {4629} (\bibinfo {year}
  {2021})}\BibitemShut {NoStop}%
\bibitem [{\citenamefont {Dion}\ \emph {et~al.}(2024)\citenamefont {Dion},
  \citenamefont {Stenning}, \citenamefont {Vanstone}, \citenamefont {Holder},
  \citenamefont {Sultana}, \citenamefont {Alatteili}, \citenamefont {Martinez},
  \citenamefont {Kaffash}, \citenamefont {T.}, \citenamefont {Kurebayashi},
  \citenamefont {Branford}, \citenamefont {Iacocca}, \citenamefont
  {Jungfleisch},\ and\ \citenamefont {Gartside}}]{Dion2024}%
  \BibitemOpen
  \bibfield  {author} {\bibinfo {author} {\bibfnamefont {T.}~\bibnamefont
  {Dion}}, \bibinfo {author} {\bibfnamefont {K.~D.}\ \bibnamefont {Stenning}},
  \bibinfo {author} {\bibfnamefont {A.}~\bibnamefont {Vanstone}}, \bibinfo
  {author} {\bibfnamefont {H.~H.}\ \bibnamefont {Holder}}, \bibinfo {author}
  {\bibfnamefont {R.}~\bibnamefont {Sultana}}, \bibinfo {author} {\bibfnamefont
  {G.}~\bibnamefont {Alatteili}}, \bibinfo {author} {\bibfnamefont
  {V.}~\bibnamefont {Martinez}}, \bibinfo {author} {\bibfnamefont {M.~T.}\
  \bibnamefont {Kaffash}}, \bibinfo {author} {\bibfnamefont {K.}~\bibnamefont
  {T.}}, \bibinfo {author} {\bibfnamefont {H.}~\bibnamefont {Kurebayashi}},
  \bibinfo {author} {\bibfnamefont {W.~R.}\ \bibnamefont {Branford}}, \bibinfo
  {author} {\bibfnamefont {E.}~\bibnamefont {Iacocca}}, \bibinfo {author}
  {\bibfnamefont {B.~M.}\ \bibnamefont {Jungfleisch}},\ and\ \bibinfo {author}
  {\bibfnamefont {J.~C.}\ \bibnamefont {Gartside}},\ }\bibfield  {title}
  {\bibinfo {title} {Ultrastrong magnon-magnon coupling and chiral spin texture
  control in a dipolar 3d multilayered artificial spin-vortex ice},\ }\href
  {https://doi.org/10.1038/s41467-024-48080-z} {\bibfield  {journal} {\bibinfo
  {journal} {Nature Communications}\ }\textbf {\bibinfo {volume} {15}},\
  \bibinfo {pages} {4077} (\bibinfo {year} {2024})}\BibitemShut {NoStop}%
\bibitem [{\citenamefont {Berchialla}\ \emph {et~al.}(2024)\citenamefont
  {Berchialla}, \citenamefont {Macauley},\ and\ \citenamefont
  {Heyderman}}]{Berchialla2024}%
  \BibitemOpen
  \bibfield  {author} {\bibinfo {author} {\bibfnamefont {L.}~\bibnamefont
  {Berchialla}}, \bibinfo {author} {\bibfnamefont {G.~M.}\ \bibnamefont
  {Macauley}},\ and\ \bibinfo {author} {\bibfnamefont {L.~J.}\ \bibnamefont
  {Heyderman}},\ }\bibfield  {title} {\bibinfo {title} {Focus on
  three-dimensional artificial spin ice},\ }\href
  {https://doi.org/10.1063/5.0229120} {\bibfield  {journal} {\bibinfo
  {journal} {Applied Physics Letters}\ }\textbf {\bibinfo {volume} {125}},\
  \bibinfo {pages} {220501} (\bibinfo {year} {2024})},\ \Eprint
  {https://arxiv.org/abs/https://pubs.aip.org/aip/apl/article-pdf/doi/10.1063/5.0229120/20269806/220501\_1\_5.0229120.pdf}
  {https://pubs.aip.org/aip/apl/article-pdf/doi/10.1063/5.0229120/20269806/220501\_1\_5.0229120.pdf}
  \BibitemShut {NoStop}%
\bibitem [{\citenamefont {Olive}\ and\ \citenamefont
  {Molho}(1998)}]{Olive1998}%
  \BibitemOpen
  \bibfield  {author} {\bibinfo {author} {\bibfnamefont {E.}~\bibnamefont
  {Olive}}\ and\ \bibinfo {author} {\bibfnamefont {P.}~\bibnamefont {Molho}},\
  }\bibfield  {title} {\bibinfo {title} {Thermodynamic study of a lattice of
  compass needles in dipolar interaction},\ }\href
  {https://doi.org/10.1103/PhysRevB.58.9238} {\bibfield  {journal} {\bibinfo
  {journal} {Phys. Rev. B}\ }\textbf {\bibinfo {volume} {58}},\ \bibinfo
  {pages} {9238} (\bibinfo {year} {1998})}\BibitemShut {NoStop}%
\bibitem [{\citenamefont {Mellado}\ \emph {et~al.}(2012)\citenamefont
  {Mellado}, \citenamefont {Concha},\ and\ \citenamefont
  {Mahadevan}}]{Mellado2012}%
  \BibitemOpen
  \bibfield  {author} {\bibinfo {author} {\bibfnamefont {P.}~\bibnamefont
  {Mellado}}, \bibinfo {author} {\bibfnamefont {A.}~\bibnamefont {Concha}},\
  and\ \bibinfo {author} {\bibfnamefont {L.}~\bibnamefont {Mahadevan}},\
  }\bibfield  {title} {\bibinfo {title} {Macroscopic magnetic frustration},\
  }\href {https://doi.org/10.1103/PhysRevLett.109.257203} {\bibfield  {journal}
  {\bibinfo  {journal} {Phys. Rev. Lett.}\ }\textbf {\bibinfo {volume} {109}},\
  \bibinfo {pages} {257203} (\bibinfo {year} {2012})}\BibitemShut {NoStop}%
\bibitem [{\citenamefont {Teixeira}\ \emph {et~al.}(2024)\citenamefont
  {Teixeira}, \citenamefont {Bernardo}, \citenamefont {Nascimento},
  \citenamefont {Saccone}, \citenamefont {Caravelli}, \citenamefont {Nisoli},\
  and\ \citenamefont {{de Araujo}}}]{Teixeira2024}%
  \BibitemOpen
  \bibfield  {author} {\bibinfo {author} {\bibfnamefont {H.}~\bibnamefont
  {Teixeira}}, \bibinfo {author} {\bibfnamefont {M.}~\bibnamefont {Bernardo}},
  \bibinfo {author} {\bibfnamefont {F.}~\bibnamefont {Nascimento}}, \bibinfo
  {author} {\bibfnamefont {M.}~\bibnamefont {Saccone}}, \bibinfo {author}
  {\bibfnamefont {F.}~\bibnamefont {Caravelli}}, \bibinfo {author}
  {\bibfnamefont {C.}~\bibnamefont {Nisoli}},\ and\ \bibinfo {author}
  {\bibfnamefont {C.}~\bibnamefont {{de Araujo}}},\ }\bibfield  {title}
  {\bibinfo {title} {Macroscopic magnetic monopoles in a 3d-printed
  mechano-magnet},\ }\href
  {https://doi.org/https://doi.org/10.1016/j.jmmm.2024.171929} {\bibfield
  {journal} {\bibinfo  {journal} {Journal of Magnetism and Magnetic Materials}\
  }\textbf {\bibinfo {volume} {596}},\ \bibinfo {pages} {171929} (\bibinfo
  {year} {2024})}\BibitemShut {NoStop}%
\bibitem [{\citenamefont {Peroor}\ \emph {et~al.}()\citenamefont {Peroor},
  \citenamefont {Scafuri}, \citenamefont {Bozhko},\ and\ \citenamefont
  {Iacocca}}]{Peroor2024}%
  \BibitemOpen
  \bibfield  {author} {\bibinfo {author} {\bibfnamefont {R.}~\bibnamefont
  {Peroor}}, \bibinfo {author} {\bibfnamefont {L.}~\bibnamefont {Scafuri}},
  \bibinfo {author} {\bibfnamefont {D.~A.}\ \bibnamefont {Bozhko}},\ and\
  \bibinfo {author} {\bibfnamefont {E.}~\bibnamefont {Iacocca}},\ }\bibfield
  {title} {\bibinfo {title} {Frequency comb in a macroscopic mechano-magnetic
  artificial spin ice},\ }\href {https://arxiv.org/abs/2409.13658} {\ ,\
  \bibinfo {pages} {arXiv:2409.13658}}\BibitemShut {NoStop}%
\bibitem [{\citenamefont {Gliga}\ \emph {et~al.}(2013)\citenamefont {Gliga},
  \citenamefont {K\'akay}, \citenamefont {Hertel},\ and\ \citenamefont
  {Heinonen}}]{Gliga2013}%
  \BibitemOpen
  \bibfield  {author} {\bibinfo {author} {\bibfnamefont {S.}~\bibnamefont
  {Gliga}}, \bibinfo {author} {\bibfnamefont {A.}~\bibnamefont {K\'akay}},
  \bibinfo {author} {\bibfnamefont {R.}~\bibnamefont {Hertel}},\ and\ \bibinfo
  {author} {\bibfnamefont {O.~G.}\ \bibnamefont {Heinonen}},\ }\bibfield
  {title} {\bibinfo {title} {Spectral analysis of topological defects in an
  artificial spin-ice lattice},\ }\href
  {https://doi.org/10.1103/PhysRevLett.110.117205} {\bibfield  {journal}
  {\bibinfo  {journal} {Phys. Rev. Lett.}\ }\textbf {\bibinfo {volume} {110}},\
  \bibinfo {pages} {117205} (\bibinfo {year} {2013})}\BibitemShut {NoStop}%
\bibitem [{\citenamefont {Venkat}\ \emph {et~al.}(2013)\citenamefont {Venkat},
  \citenamefont {Kumar}, \citenamefont {Franchin}, \citenamefont {Dmytriiev},
  \citenamefont {Mruczkiewicz}, \citenamefont {Fangohr}, \citenamefont
  {Barman}, \citenamefont {Krawczyk},\ and\ \citenamefont
  {Prabhakar}}]{Venkat2013}%
  \BibitemOpen
  \bibfield  {author} {\bibinfo {author} {\bibfnamefont {G.}~\bibnamefont
  {Venkat}}, \bibinfo {author} {\bibfnamefont {D.}~\bibnamefont {Kumar}},
  \bibinfo {author} {\bibfnamefont {M.}~\bibnamefont {Franchin}}, \bibinfo
  {author} {\bibfnamefont {O.}~\bibnamefont {Dmytriiev}}, \bibinfo {author}
  {\bibfnamefont {M.}~\bibnamefont {Mruczkiewicz}}, \bibinfo {author}
  {\bibfnamefont {H.}~\bibnamefont {Fangohr}}, \bibinfo {author} {\bibfnamefont
  {A.}~\bibnamefont {Barman}}, \bibinfo {author} {\bibfnamefont
  {M.}~\bibnamefont {Krawczyk}},\ and\ \bibinfo {author} {\bibfnamefont
  {A.}~\bibnamefont {Prabhakar}},\ }\bibfield  {title} {\bibinfo {title}
  {Proposal for a standard micromagnetic problem: spin wave dispersion in a
  magnonic waveguide},\ }\href {https://ieeexplore.ieee.org/document/6228538}
  {\bibfield  {journal} {\bibinfo  {journal} {IEEE Trans. Magn.}\ }\textbf
  {\bibinfo {volume} {49}},\ \bibinfo {pages} {524} (\bibinfo {year}
  {2013})}\BibitemShut {NoStop}%
\bibitem [{\citenamefont {Wagner}\ \emph {et~al.}(2016)\citenamefont {Wagner},
  \citenamefont {K\'{a}kay}, \citenamefont {Schultheiss}, \citenamefont
  {Henschke}, \citenamefont {Sebastian},\ and\ \citenamefont
  {Schultheiss}}]{Wagner2016}%
  \BibitemOpen
  \bibfield  {author} {\bibinfo {author} {\bibfnamefont {K.}~\bibnamefont
  {Wagner}}, \bibinfo {author} {\bibfnamefont {A.}~\bibnamefont {K\'{a}kay}},
  \bibinfo {author} {\bibfnamefont {K.}~\bibnamefont {Schultheiss}}, \bibinfo
  {author} {\bibfnamefont {A.}~\bibnamefont {Henschke}}, \bibinfo {author}
  {\bibfnamefont {T.}~\bibnamefont {Sebastian}},\ and\ \bibinfo {author}
  {\bibfnamefont {H.}~\bibnamefont {Schultheiss}},\ }\bibfield  {title}
  {\bibinfo {title} {Magnetic domain walls as reconfigurable spin-wave
  nanochannels},\ }\href@noop {} {\bibfield  {journal} {\bibinfo  {journal}
  {Nat. Nanotechnol.}\ }\textbf {\bibinfo {volume} {11}},\ \bibinfo {pages}
  {432} (\bibinfo {year} {2016})}\BibitemShut {NoStop}%
\end{thebibliography}
\end{document}